\documentclass[a4paper]{iopart}
\usepackage{iopams}     % IOP version of amsmath and amssymb packages
\usepackage{graphicx}
\usepackage{epsfig}
\usepackage{color}
\usepackage{url}
\usepackage{times}
\usepackage{bm}         % bold math symbols

\usepackage[utf8]{inputenc}
\usepackage[english,american]{babel}
\usepackage{latexsym}

\definecolor{darkred}{rgb}{0.5,0,0}
\definecolor{darkgreen}{rgb}{0,0.5,0}
\definecolor{darkblue}{rgb}{0,0,0.5}
\definecolor{black}{rgb}{0,0,0}

\newcommand{\Msun}{\,M_{\odot}}

\newcommand{\bea}{\begin{eqnarray}}
\newcommand{\eea}{\end{eqnarray}}
\newcommand{\beq}{\begin{equation}}
\newcommand{\eeq}{\end{equation}}

\newcommand{\cf}{\textit{cf.}~}
\newcommand{\ie}{\textit{i.e.}~}
\newcommand{\eg}{\textit{e.g.}~}
\newcommand{\mss}{{\rm ms}}

\usepackage{ifpdf}
\usepackage{graphicx}
\newcommand{\imgname}[1]{#1.eps}

\usepackage{ulem}

\definecolor{magenta}{cmyk}{0.1,0.8,0,0.1} 

\definecolor{darkred}{rgb}{0.5,0,0}

\newcommand{\apj}{Astrophys. J.}
\newcommand{\dr}{{\mathrm d}}
\newcommand{\aap}{Astronomy and Astrophysics}
\newcommand{\prd}{Physical Review D}

\ifpdf  \fi \ifpdf
\renewcommand{\imgname}[1]{#1.pdf}
 \fi
\begin{document}

\title[On the Shear Instability in Relativistic Neutron Stars]{
On the Shear Instability in Relativistic Neutron Stars}

\author[G.~Corvino et al.]{Giovanni Corvino$^{1}$, Luciano Rezzolla$^{1}$, 
        Sebastiano Bernuzzi$^{2}$, Roberto De Pietri$^{3}$
        and Bruno Giacomazzo$^{4,5,1}$ 
}

\address{$^1$ Max-Planck-Institut f\"ur Gravitationsphysik,
  Albert-Einstein-Institut, Golm, Germany}

\address{$^2$ Theoretical Physics Institute, University of Jena, 07743
  Jena, Germany }

\address{$^3$ Physics Department, Parma University and INFN, Parma, Italy}

\address{$^4$ Department of Astronomy, University of Maryland, College
  Park, MD, USA}

\address{$^5$ Gravitational Astrophysics Laboratory, NASA Goddard
  Space Flight Center, Greenbelt, MD, USA}

\date{\today}
\begin{abstract}
  We present new results on instabilities in rapidly and
  differentially rotating neutron stars. We model the stars in full
  general relativity and describe the stellar matter adopting a cold
  realistic equation of state based on the unified SLy
  prescription~\cite{Douchin01}. We provide evidence that rapidly and
  differentially rotating stars that are below the expected threshold
  for the dynamical bar-mode instability, $\beta_c \equiv T/|W| \simeq
  0.25$, do nevertheless develop a shear instability on a dynamical
  timescale and for a wide range of values of $\beta$. This class of
  instability, which has so far been found only for small values of
  $\beta$ and with very small growth rates, is therefore more generic
  than previously found and potentially more effective in producing
  strong sources of gravitational waves. Overall, our findings support
  the phenomenological predictions made by Watts, Andersson and
  Jones~\cite{Watts:2003nn} on the nature of the low-$T/|W|$
  instability as the manifestation of a shear instability in a region
  where the latter is possible only for small values of the
  $\beta$. Furthermore, our results provide additional insight on
  shear instabilities and on the necessary conditions for their
  development.
\end{abstract}

%\maketitle

%-----------------------------------------------------------------
\section{Introduction}
\label{sec:intro}
%-----------------------------------------------------------------

Non-axisymmetric deformations of rapidly rotating bodies are rather
generic phenomena in Nature and can appear in a wide class of
systems. Particularly interesting within an astrophysical context are
those deformations taking place in fluids that are self-gravitating
and the literature on this has a long history dating back to the work
of~\cite{Chandrasekhar69c} on incompressible Newtonian uniformly
rotating bodies. Since then, the study of these instabilities has
continued over the years both in Newtonian gravity and in full general
relativity.

Special attention has traditionally been paid to the study of $m=2$
instabilities, which are characterized by the exponential growth of
$m=2$ deformations, where $m$ parametrizes the azimuthal dependence
${\rm e}^{{\rm i} m\phi}$ in a standard mode decomposition in spherical
harmonics. Most of the interest in this type of deformation in compact
stars stems from the fact that it has the shortest growth time and
leads to the emission of a strong gravitational-wave signal.

The development of non-axisymmetric instabilities is commonly analyzed
in terms of the quantity $\beta\equiv T/|W|$ (\ie the ratio between
the kinetic rotational energy $T$ and the gravitational potential
energy $W$), that provides a dimensionless measure of the amount of
angular momentum that can be tapped to feed the development of the
instabilities.  This parameter plays an important role in what is
possibly the most celebrated of the non-axisymmetric instabilities:
the so-called \textit{dynamical} bar-mode instability. This is an
$m=2$ instability which takes place when the parameter $\beta$ is
larger than a critical one, $\beta_c$. In the case of a Newtonian
incompressible self-gravitating polytrope, for instance, the dynamical
bar-mode instability develops for $\beta \geq
\beta_{d}=0.2738$~\cite{Chandrasekhar69c} and is only weakly dependent
on the considered polytropic index or whether the fluid is
compressible. Post-Newtonian (PN) studies~\cite{Saijo:2000qt} or fully
general-relativistic ones~\cite{Shibata:2000jt} correct this results
only slightly, by reducing the threshold to somewhat lower values of
the instability parameter. As an example, for a polytropic
relativistic star with polytropic index $\Gamma=2$, the accurate
calculations reported in~\cite{Baiotti06b} reveal that the critical
value is $\beta_c\sim 0.245$ and that a simple dependence on the
stellar compactness allows one to track this threshold from the
Newtonian limit over to the fully relativistic one~\cite{Manca07}.

The onset and development of the bar-mode instability has been
traditionally studied by means of nonlinear 3D simulations of
Newtonian stars that are either unmagnetized~\cite{Brown2000,
  Houser96, Liu02, New2000, Pickett96} and, more recently, also
magnetized~\cite{Camarda:2009mk}. In addition, PN and fully
relativistic simulations have been performed and highlighted, for
instance, that the persistence of the bar is strongly dependent on the
degree of overcriticality and is generically of the order of the
dynamical timescale. Furthermore, generic nonlinear mode-coupling
effects between the $m=1$ and the $m=2$ mode appear during the
development of the instability and these can severely limit the
persistence of the bar deformation and eventually suppress the bar
deformation~\cite{Baiotti06b}. These results have been recently
confirmed by the perturbative calculations in~\cite{Saijo2008}.

Besides dynamical instabilities, which are purely hydrodynamical,
\textit{secular} instabilities are also possible in rotating compact
stars and these are instead triggered by dissipative processes, such
as viscosity or radiation emission. If, in particular, the dissipative
mechanism is the emission of gravitational radiation, then the secular
instability is also known as Chandrasekhar-Friedman-Schutz or CFS
instability~\cite{Chandrasekhar:1970,Friedman78}. Contrary to what
their name may suggest, secular instabilities do not necessarily
develop on secular timescales (although they normally do) and are
characterized by having a much smaller threshold for the
instability. Once again, in the case of a Newtonian polytrope, the
critical secular instability parameter is as small as $\beta_c\sim
0.14$ and thus much more easy to attain in astrophysical
circumstances.
 
Although widely observed in numerical simulations, the physical
conditions leading to a dynamical bar-mode instability are difficult
to be encountered in standard astrophysical scenarios. Such large
values of the instability parameter, in fact, cannot be easily
attained in old and cold neutron stars, which have been brought into
uniform rotation and thus to rather small values of $\beta$. However,
more recently these pessimistic prospects have been changed when a new
$m=2$ instability has been discovered in differentially rotating
Newtonian stars~\cite{Shibata:2002mr} for values of $\beta\approx
0.01$, therefore well below the expected values for a dynamical
bar-mode instability. The most salient aspect of this new instability
is that it appears in stars with a large degree of differential
rotation and that it grows on a timescale which is longer but
comparable with the dynamical one. This instability has been referred
to as the ``low-$T/|W|$ instability'' and its dependence on the
polytropic index and on the degree of differential rotation has been
studied in~\cite{Shibata:2003yj}. Since then, the instability has been
observed or discussed in a number of related studies~\cite{Ott:2005gj,
  Ou06, Cerda07b, Ott07a, Ott07b,scheidegger08,Abdikamalov:2009aq},
all of which have highlighted the possible occurrence of this type of
instability during the collapse of a massive stellar core.

Despite the abundant numerical evidence on the development of this
instability, the nature of these low-$T/|W|$ instabilities is still
matter of debate and, most importantly, a sufficient criterion for its
onset has not been derived yet. This instability has been studied in
great detail by Watts and
collaborators~\cite{Watts:2002ik,Watts:2003nn}, who have made a number
of phenomenological predictions either using a toy shell-model first
introduced in~\cite{Rezzolla:2001rh_a, Rezzolla:2001rh_b}, or for a
stellar model in Newtonian gravity. Overall, the work of Watts and
collaborators (but see also~\cite{Saijo:2005gb}) recognizes the
low-$T/|W|$ instabilities as the manifestation of a more generic class
of instabilities, the \textit{shear instabilities}~\cite{Watts:2003nn}
, that is unstable oscillations that do not exist in uniformly
rotating systems and are associated to the existence of a corotation
band~\cite{Balbinski85b,Luyten:1990}. Watts, Andersson and Jones
suggest, in particular, that a necessary condition for the development
of the instability is ``corotation'', that is the presence of a point
at which the star rotates at the same pattern speed of the unstable
mode~\cite{Watts:2003nn}. An alternative suggestion on the necessary
conditions has been made also by Ou and Tohline~\cite{Ou06}, who
instead associate the development of the instability to the presence
of a minimum in the vortensity profile of the star. This minimum can
then drive unstable not only the corotating $m=2$-modes but also the
odd modes such as the $m=1$ and $m=3$-modes~\cite{Papaloizou89}. In
this interpretation, the growth time of the instability is
proportional to distance between the corotation radius, \ie the radial
position at which the unstable mode corotates with the star, and the
minimum of the vortensity.
 
The purpose of this work is to shed some light on the development of
shear instabilities and, in particular, to validate one prediction
made, although not explicitly, by Watts, Andersson and Jones. More
specifically, we show that, for sufficient amounts of differential
rotation, shear instabilities develop for \textit{any value} of the
instability parameter $\beta$ and also below the expected critical
value for the dynamical bar-mode instability. We therefore provide
evidence that the low-$T/|W|$ instability is not a new instability but
rather the manifestation of a shear instability in a region where the
latter is possible only for small values of $\beta$.

Our analysis proceeds via the simulation in full general relativity of
sequences of neutron star models having constant rest-mass and
constant degrees of differential rotation, but with different amounts
of rotation, \ie with different values of $\beta$. The neutron-star
matter is described by a realistic equation of state (EOS) defined by
the unified SLy prescription~\cite{Douchin01} and we study the
development of the non-axisymmetric instabilities from their linear
growth up to the fully nonlinear development and suppression.
Interestingly, we find that depending on the degree of differential
rotation, the shear instability leads either to the growth of a single
modes (for the low-$\beta$ models) or to the simultaneous presence of
up to three unstable modes (for the high-$\beta$ models), which
produce beatings in the growth of the overall $m=2$ deformation.
Special attention is also paid to the properties of the unstable modes
and to their position within the corotation band or the vortensity
profiles. In this way we are able to confirm both the necessary
conditions proposed so far for the onset of the instability.  In
particular, we show that all the unstable modes are within the
corotation band of the progenitor axisymmetric model~(\cf
\cite{Watts:2003nn}) and that all of the unstable models have
vortensity profiles with a local minimum (\cf ~\cite{Ou06}).

The structure of the paper is as follows: in section
\ref{sec:numerics} we describe the numerical setting of our
simulations, the EOS we used and the initial models we generated. In
section \ref{sec:methodology} we describe the quantities and tools we
used to monitor the evolution of the instability.  In section
\ref{sec:results} we report the results of the simulations and section
\ref{sec:conclusions} is dedicated to conclusions and discussion.  We
use a spacelike signature $(-,+,+,+)$ and a system of units in which
$c=G=\Msun=1$ (or in cgs units whenever more convenient). Greek
indices are taken to run from $0$ to $3$, Latin indices from $1$ to
$3$ and we adopt the standard convention for the summation over
repeated indices.

%XXXXXXXXXXXXXXXXXXXXXXXXXXXXXXXXXXXXXXXXXXXXXXXXXXXXXXXXXXXXXXXXXXXXXXXXXX%
%-----------------------------------------------------------------
\section{Numerical Setup and Initial Models}
\label{sec:numerics}
%-----------------------------------------------------------------

In what follows we provide a brief overview of the numerical setup
used in the simulations, of the realistic EOS adopted and on the
procedure followed for the construction of the initial axisymmetric
models.

%-----------------------------------------------------------------
\subsection{Numerical Setup}
\label{sec:eqs}
%-----------------------------------------------------------------

We solve numerically the full set of Einstein equations
\begin{equation}
\label{eq:Einstein}
G_{\mu\nu} = 8 \pi T_{\mu\nu} \,,
\end{equation}
where $G_{\mu\nu}$ and $T_{\mu\nu}$ are the Einstein tensor and
the stress-energy tensor, respectively. The equations are solved
within the ``3+1'' decomposition of spacetime, in which the
$4$-dimensional metric $g_{\mu\nu}$ is decomposed into the spatial
metric $\gamma_{ij}$, the lapse function $\alpha$ and the shift vector
components $\beta_i$. The field equations, which then also provide an
evolution for the extrinsic curvature tensor $K_{ij}$, are then
coupled to those of general relativistic hydrodynamics
\begin{equation}
\label{eq:conservation}
\nabla_{\mu} T^{\mu\nu}= 0 \;\; ; \;\; \nabla_{\mu} (\rho u^{\mu}) = 0 \ ,
\end{equation}
where, in the case of a perfect-fluid, the stress-energy tensor is
given by
\begin{equation}
\label{eq:MATTER}
T^{\mu\nu} = \rho \left(1 +\epsilon + \frac{p}{\rho} \right) 
              u^{\mu} u^{\nu} + p g^{\mu\nu} \, .
\end{equation}
Above $u^\mu$ is the fluid $4$-velocity, $p$ is the fluid pressure,
$\epsilon$ the specific internal energy and $\rho$ the rest-mass
density, so that $e = \rho (1+\epsilon)$ is the energy density in the
rest frame of the fluid. The set of hydrodynamics equation is then
closed by a prescription for the properties of the matter in the form
of a relation between the pressure and other quantities in the fluid,
\eg $p=P(\rho,\epsilon)$, and for which we have chosen a cold and
realistic EOS which will be discussed in the following section.

The evolution of the spacetime was performed using the \texttt{CCATIE}
code, a three-dimensional finite-differencing code providing a
solution of a conformal traceless formulation of the Einstein
equations (see~\cite{Pollney:2007ss} for the explicit expressions of
the equations solved in the code and also~\cite{Moesta:2009} for a
more recent and improved implementation). The relativistic
hydrodynamics equations, on the other hand, were solved using the
\texttt{Whisky} code, which adopts a flux-conservative formulation of
the equations as presented in~\cite{Banyuls97} and high-resolution
shock-capturing schemes or HRSC
(see~\cite{Baiotti03a,Baiotti04,Baiotti07} for the explicit
expressions of the equations solved in the code and
also~\cite{Giacomazzo:2009mp} for a more recent extension of the code
to MHD). The \texttt{Whisky} code implements several reconstruction
methods, such as Total-Variation-Diminishing (TVD) methods,
Essentially-Non-Oscillatory (ENO) methods~\cite{Harten87} and the
Piecewise Parabolic Method (PPM)~\cite{Colella84}. Also, a variety of
approximate Riemann solvers can be used, starting from the
Harten-Lax-van Leer-Einfeldt (HLLE) solver~\cite{Harten83}, over to
the Roe solver~\cite{Roe81} and the Marquina flux
formula~\cite{Aloy99b} (see~\cite{Baiotti03a,Baiotti04} for a more
detailed discussion). All the results reported hereafter have been
computed using the Marquina flux formula and a PPM reconstruction.

Both the Einstein and the hydrodynamics equations are solved using the
vertex-centered adaptive mesh-refinement (AMR) approach provided by
the \texttt{Carpet} driver~\cite{Schnetter-etal-03b}. Our rather basic
form of AMR consists of box-in-box structures centered on the origin
of the coordinate system and with the finest grid covering the whole
star at all times. The simulations reported here make use of $4$
levels of refinement, with the finest having a resolution of $221\,
\mathrm{m}$ and the coarsest one a resolution of
$1.77\,\mathrm{km}$. The outer boundary was set relatively close to
the star and at a distance of $\simeq 159.5\ \mathrm{km}$, \ie at
about $\simeq 10$ times the size of the star. A reflection symmetry
across the $(x,y)$ (equatorial symmetry) plane was used to reduce the
computational costs, but not a rotational one around the $z$-axis
($\pi$-symmetry) as it would have artificially prevented the growth of
odd-$m$ modes (see discussion in~\cite{Baiotti06b}).

%-----------------------------------------------------------------
\subsection{Realistic Equation of State}
\label{sec:EOS}
%-----------------------------------------------------------------

As mentioned above, the system of hydrodynamics equations needs to be
closed by an EOS relating the pressure with the other primitive
variables, \eg the rest-mass density. Previous studies of the bar-mode
instability, both in Newtonian gravity and in general relativity, have
been focused on the use of ideal fluids and analytic EOSs, either in
the form a of a ``polytropic'' (and isentropic) EOS $p=p(\rho)$, or of
an ``ideal-fluid'' and (non-isentropic) EOS $p=p(\rho,\epsilon)$ (\cf
discussion in~\cite{Baiotti06b,Manca07}). While these two descriptions
are expected to provide results that are qualitatively correct, a more
accurate modelling of these instabilities in compact stars necessarily
requires a more physically-motivated description of the neutron-star
matter.

It is in this spirit that we have here considered a realistic EOS,
namely the unified SLy EOS~\cite{Douchin01}, which models high-density
and cold (\ie zero temperature) matter via a Skyrme effective
potential for the nucleon-nucleon interactions. The SLy EOS, which
describes via a single effective Hamiltonian the neutron star's
interior, is supplemented with the HP94 EOS~\cite{Haensel94} to
describe the crustal matter and with the BPS EOS~\cite{Baym71b} for
lower density regions. This prescription results in a one-parameter
EOS in the form $p=p(\e(\rho))=p(\rho)$, where the SLy EOS is used for
$\rho > 4.979\times10^{10} g cm^{-3}$, the HP94 EOS is used for $10^{8} 
g cm^{-3}\lesssim\rho \leqslant 4.979\times10^{10} g cm^{-3}$ and 
the BPS EOS for $\rho\lesssim10^{8} g cm^{-3}$ 
(see also figure~1 of~\cite{Haensel04}). In
addition, at even lower densities the EOS becomes temperature
dependent (and thus no longer a simple barotropic EOS), but because
these these regions are well below the threshold for the artificial
atmosphere, we do not consider an additional prescription for
$\rho\lesssim10^{8}\,{\rm g\ cm}^{-3}$.

We recall, in fact, that our HRSC methods require the use of a tenuous
atmosphere which fills the regions of the computational domain not
occupied by the compact star. The threshold value for the rest-mass
density of the atmosphere is chosen to be several orders of magnitude
smaller than the maximum value and in our simulation a fluid element
is considered to be part of the atmosphere if its rest-mass density $\rho$ 
satisfies $\rho/ \max\left(\rho(t=0)\right) \leq \times 10^{-8}$. When this
happens the fluid element is treated as a non-dynamical cold fluid
described by a polytropic EOS, $p(\rho) = K\rho^\Gamma$, with
$\Gamma=2$ and $K=100$ and its velocity is set to zero (see~\cite{baiotti08} for a
more detailed discussion on the use of the atmosphere in the
\texttt{Whisky} code).

The practical implementation of the realistic EOS can take place in a
number of different ways. The simplest is to use standard
interpolation techniques, \eg based on Lagrangian polynomials, on the
values of the published tables. While straightforward, the
interpolations in this approach do not guarantee in general that the
thermodynamics relations are fulfilled (see~\cite{Swesty95a} for a
thermodynamical preserving interpolation). In addition, the
derivatives of the fields, \eg of the pressure to evaluate the sound
speed, are typically not available in the tables. Furthermore, the use
of high-order interpolation and/or finite differences can lead to
undesirable spurious oscillations.

A second approach that removes all of these problems, uses analytic
fits that have been proposed for the pressure. As an example,
ref.~\cite{Haensel04} suggested to fit the specific internal energy of
the unified SLy EOS table with the expression\footnote{Note that a
  different notation is used in~\cite{Haensel04} for some primitive
  variables.}
\begin{eqnarray}
\label{eq:analfit}
\fl
  \epsilon=\frac{p_{1}\rho^{p_{2}}+p_{3}\rho^{p_{4}}}{(1+p_{5}\rho)^{2}}f_{0}\{-p_{6}(\log(\rho)+p_{7})\} + 
  \frac{\rho}{8\times10^{-6}+2.1\rho^{0.585}}f_{0}\{p_{6}(\log(\rho)+p_{7})\}\ ,
\end{eqnarray}
where $f_0\{x\} = {1}/({e^{x}+1})$, $\rho$ and $\epsilon$ are in cgs
units, and the coefficients $p_{i}$ are
$p_{i}=\{0.320,\ 2.17,\ 0.173,\ 3.01,\ 0.540,\ 0.847,\ 3.581\}$ (see
Table~$2$ of~\cite{Haensel04}). Equation~(\ref{eq:analfit}) is
obtained from Eq.~(15) of~\cite{Haensel04} after using $\rho=m_{\rm B}
n$, where $n$ is the baryon number density and $m_{\rm
  B}=1.66\times10^{-24}\,{\rm g}$ is the mass of the
nucleons. As discussed in~\cite{Shibata05c}, it is then possible to
compute the pressure from the value of $\epsilon$ using the first
principle of thermodynamics at $T=0$
\begin{equation}
  p=\rho^{2}\frac{d\epsilon}{d\rho} \ , \label{Shibatapress}
\end{equation}
and thus to have an evaluation of the pressure which is
thermodynamically consistent. The differences between the fit and the
table are typically less then $2\%$. Unfortunately, although
apparently very convenient, the evaluation of the fitting formulas
containing several exponential and logarithmic functions, turns out to
be computationally rather expensive even if done in a optimized way.

As a third approach, which combines the efficiency of a table search
with the thermodynamical consistency of an analytic fit, consists of
performing a simple linear interpolation among the tabulated values
constructed from the analytic fit. Besides being highly efficient, a
linear interpolation also eliminates the spurious oscillations that
arise, for instance, in the derivative of the pressure if high-order
interpolation formulas are used. In this case, the interpolation error
can be reduced simply by populating the analytically constructed
tables with a large number of entries, \eg $\sim 600$ in place of the
$\sim 150$ which are typically available in published tables. This
third approach is the one actually implemented in \texttt{Whisky} and
provides a speed up of about $20\%$ with respect to the evaluation of
the pressure via the analytic fits and with comparable accuracy.

%-----------------------------------------------------------------
\subsection{Initial Data}
\label{sec:initial_data}
%-----------------------------------------------------------------

The initial data for our simulations are prepared as stationary and
axisymmetric equilibrium solutions for rapidly rotating relativistic
stars~\cite{Stergioulas95}. Adopting spherical quasi-isotropic
coordinates, the line element of the corresponding spacetime is
\begin{equation} 
  \dr s^{2} = - \e^{\mu+\nu} \dr t^{2}
  + \e^{\mu-\nu} r^{2} \sin^{2}\theta (\dr \phi-\omega \dr t)^{2}
  +\e^{2\xi}(\dr r^{2}+r^{2} \dr \theta^{2})\,,
\end{equation} 
where $\mu$, $\nu$, $\omega$ and $\xi$ are functions of $r$ and
$\theta$. Moreover we assume the usual relativistic $j$-constant law
of differential rotation and that amounts to assume an
angular-velocity distribution of the form
\begin{equation} 
  \Omega_{c} -\Omega   =   \frac{1}{\hat{A}^{2} r_e^2}
  \left[ \frac{(\Omega-\omega) r^2   \sin^2\theta \e^{-2\nu}
    }{1-(\Omega-\omega)^{2} r^2 \sin^2\theta \e^{-2\nu}
    }\right] \,,
  \label{eq:velocityProfile_gr}
\end{equation} 
where $r_{e}$ is the coordinate equatorial stellar radius and the
coefficient $\hat{A}$ provides a measure of the degree of differential
rotation. Expression~(\ref{eq:velocityProfile_gr}) represents the
general-relativistic equivalent of the simpler Newtonian $j$-constant
law~\cite{Rezzolla:2001rh_a}
\begin{equation}
\Omega_c-\Omega = \frac{\Omega_c r^2\sin^2\theta}{({\hat A}^2 r^2_e + r^2\sin^2\theta)}\,. 
\label{eq:velocityProfile_newt}
\end{equation}
Clearly, $\hat{A} \rightarrow \infty$ corresponds to a star in uniform
rotation, while $\hat{A} \rightarrow 0$ corresponds to a star with
increasing degree of differential rotation. As a reference, $\hat{A} =
1$ yields a star with an angular-velocity profile which varies of a
factor $\sim 3$ between the center and the surface of the star (\cf
left panel of figure~\ref{fig:IDa}).

In practice, we have computed a very large number of initial models
using the SLy prescription for the EOS for which we have computed {\it
  baryonic} mass $M_b$, the gravitational mass $M$, the angular
momentum $J$, the rotational kinetic energy $T$ and the gravitational
binding energy $W$ defined as
\begin{eqnarray}
  &M_b \equiv \displaystyle \int\!d^3\!x\,  \sqrt{\gamma} W_{_{\!L}} \rho \,, 
  \qquad 
  &M \equiv \displaystyle \int\!d^3\!x\, \left( -2T^0_0+T^\mu_\mu\right) 
  \alpha \sqrt{\gamma}\,,
\label{eq:DEF M}
\\
&{E_{\rm int}} \equiv \displaystyle \int\!d^3\!x\,  \sqrt{\gamma} W_{_{\!L}}
\rho \epsilon\,, 
\qquad
&J \equiv  \displaystyle \int\!d^3\!x\, T^0_\phi \alpha \sqrt{\gamma}\,,
\label{eq:DEF J}
\\
&T \equiv \displaystyle \frac{1}{2} \int\!d^3\!x\, \Omega  T^0_\phi
\alpha \sqrt{\gamma}\,, 
&W\equiv  T + {E_{\rm int}} + M_b - M \,,
\label{eq:DEF W}
\end{eqnarray}
where $\sqrt{\gamma}$ is the square root of the determinant of
three-dimensional metric $\gamma_{ij}$ and $W_{_{\!L}}=\alpha u^0$ is
the fluid Lorentz factor. We stress that the definitions~(\ref{eq:DEF
  M})--(\ref{eq:DEF W}) of quantities such as $J$, $T$, $W$ and
$\beta$ are meaningful only in the case of stationary axisymmetric
configurations and should therefore be treated with care once the
rotational symmetry is lost.

\begin{figure}[t]
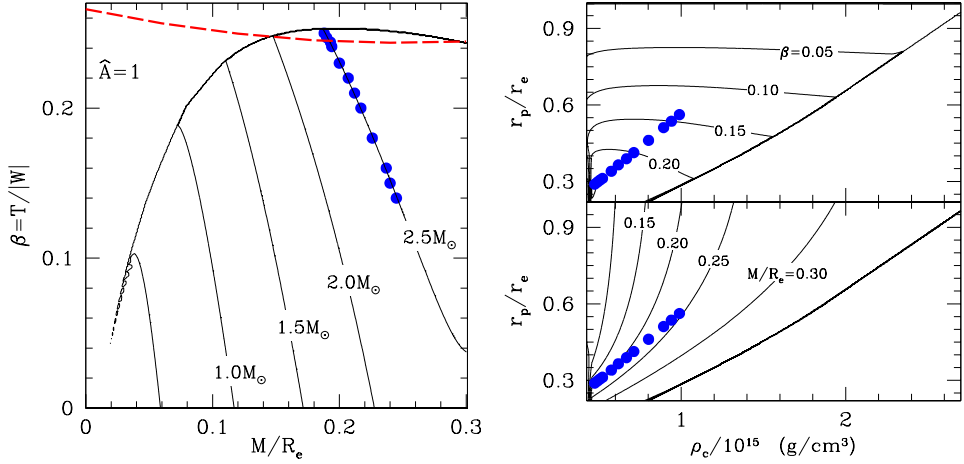

\begin{center}
\includegraphics[width=0.49\textwidth]{\imgname{ID_A.eq.1.0}}
\includegraphics[width=0.49\textwidth]{\imgname{ID2_A.eq.1.0}}
\end{center}
\vglue-.5cm
\caption{Left panel: position in the $(\beta, M/R_e)$ plane of the
  initial models computed with ${\hat A}=1$, with the filled circles
  representing those we have evolved numerically (A similar behaviour
  is shown also by the models with ${\hat A}=2$. Indicated with solid
  thin lines are isocontours of constant baryon mass models while
  indicated with a thick dashed line is the threshold to the dynamical
  bar-mode instability as computed for a $\Gamma=2$
  polytrope~\cite{Manca07}. Note that the threshold for the
  instability tends to increase for smaller rest-masses. Right panel:
  the same initial models as in the left panel but shown in $(\rho_c,
  r_p/r_e)$ planes with isocontours of constant $\beta$ (upper part)
  or constant $M/R_e$ (lower part). See Table~\ref{table:initial} for
  a summary of the properties of the initial models.}
\label{fig:IDb}
\end{figure}
%%%----------------------------------------------------------------
%%% inital model table
%%%----------------------------------------------------------------
\begin{table}
  \caption{Main properties of the simulated stellar models. Starting
    from the left: the name of the model, the differential rotation 
    parameter $\hat{A}$the instability parameter $\beta$, the 
    compactness $M/R_e$, the central rest-mass
    density $\rho_c$, the maximum of the density $\rho_{\rm max}$
    (Note that for models with ${\hat A}=2$, $\rho_{\rm
      max}=\rho_c$.), the ratio between the polar and the equatorial
    coordinate radii $r_{p}/r_{e}$, the proper equatorial radius
    $R_{e}$, the gravitational mass $M$, the total angular momentum
    $J$ divided by the square of the gravitational mass.
  } %%%%%%%%%%%%%%%%%%%%%%%%%%%%%%%%%%%%%%%%%%%%%%%%%%%%%%%%
  \label{table:initial}
  \begin{tabular}{@{}cccccccccc}
      \br
Model & $\hat{A}$ & $\beta$    & $M/R_e\!\!$& $\rho_c/10^{15}$ & $\rho_{\rm max}/10^{15}$ & $r_p/r_e$ &$R_e$ &$M$ & $J/M^2$ \\[0mm] 
& & & & (${\rm g/cm^3}$) & (${\rm g/cm^3}$) & & (Km) & $(M_{\odot})$ &  \\[0mm] 
\hline
\texttt{M.1.140} & $1$ & $0.140$ & $0.245$ & $0.989$ & $1.011$ & $0.562$ & $13.20$ & $2.19$ & $0.681$ \\ %& $0.298$ & $1.107$\\  
\texttt{M.1.150} & $1$ & $0.150$ & $0.240$ & $0.941$ & $0.968$ & $0.536$ & $13.48$ & $2.20$ & $0.708$ \\ %& $0.300$ & $1.099$\\   
\texttt{M.1.160} & $1$ & $0.160$ & $0.237$ & $0.893$ & $0.926$ & $0.511$ & $13.81$ & $2.21$ & $0.735$ \\ %& $0.303$ & $1.099$\\ 
\texttt{M.1.180} & $1$ & $0.180$ & $0.226$ & $0.802$ & $0.847$ & $0.461$ & $14.49$ & $2.22$ & $0.789$ \\ %& $0.313$ & $1.107$\\ 
\texttt{M.1.200} & $1$ & $0.200$ & $0.217$ & $0.712$ & $0.773$ & $0.413$ & $15.27$ & $2.24$ & $0.844$ \\ %& $0.327$ & $1.127$\\ 
\texttt{M.1.210} & $1$ & $0.210$ & $0.212$ & $0.668$ & $0.737$ & $0.389$ & $15.70$ & $2.25$ & $0.873$ \\ %& $0.336$ & $1.140$\\ 
\texttt{M.1.220} & $1$ & $0.220$ & $0.207$ & $0.618$ & $0.703$ & $0.365$ & $16.16$ & $2.26$ & $0.902$ \\ %& $0.347$ & $1.161$\\  
\texttt{M.1.230} & $1$ & $0.230$ & $0.200$ & $0.576$ & $0.669$ & $0.340$ & $16.67$ & $2.26$ & $0.933$ \\ %& $0.359$ & $1.184$\\
\texttt{M.1.241} & $1$ & $0.241$ & $0.194$ & $0.522$ & $0.633$ & $0.312$ & $17.29$ & $2.27$ & $0.968$ \\ %& $0.375$ & $1.216$\\ 
\texttt{M.1.244} & $1$ & $0.244$ & $0.193$ & $0.506$ & $0.624$ & $0.305$ & $17.47$ & $2.28$ & $0.978$ \\ %& $0.380$ & $1.226$\\ 
\texttt{M.1.247} & $1$ & $0.247$ & $0.190$ & $0.490$ & $0.614$ & $0.297$ & $17.66$ & $2.28$ & $0.988$ \\ %& $0.385$ & $1.236$\\ 
\texttt{M.1.250} & $1$ & $0.250$ & $0.188$ & $0.474$ & $0.604$ & $0.289$ & $17.85$ & $2.28$ & $0.999$ \\ %& $0.391$ & $1.246$\\ 
\hline
\texttt{M.2.125} & $2$ & $0.125$ & $0.241$ & $1.143$ & $1.143$ & $0.642$ & $13.22$ & $2.17$ & $0.668$ \\ %& $0.509$ & $0.902$\\
\texttt{M.2.150} & $2$ & $0.150$ & $0.227$ & $1.039$ & $1.039$ & $0.578$ & $14.15$ & $2.17$ & $0.739$ \\ %& $0.517$ & $0.896$\\
\texttt{M.2.175} & $2$ & $0.175$ & $0.210$ & $0.947$ & $0.947$ & $0.512$ & $15.33$ & $2.18$ & $0.801$ \\ %& $0.531$ & $0.900$\\
\texttt{M.2.200} & $2$ & $0.200$ & $0.184$ & $0.865$ & $0.865$ & $0.435$ & $17.34$ & $2.15$ & $0.878$ \\ %& $0.574$ & $0.941$\\
\hline
%\br
\end{tabular}
%\end{indented}
\end{table}
%%%----------------------------------------------------------------
%%%----------------------------------------------------------------

Out of this large set, we have then selected for the numerical
evolution a number of models so as to build sequences of constant
baryonic mass with $M_b=2.5\, M_\odot$, degree of differential
rotation ${\hat A}=1, 2$ and values of the instability parameter
$\beta$ ranging between $0.140$ and $0.250$. Each model in this
sequence is supramassive, namely it has a mass which is larger than
the maximum mass allowed for a corresponding nonrotating model, \ie
$M_{max}\vert_{\Omega=0}=2.05\,\Msun,
M_{b,max}\vert_{\Omega=0}=2.43\,\Msun$, although it is not
hypermassive, namely it does not have mass which is above the maximum
mass for a uniformly rotating model, \ie
$M_{max}\vert_{\Omega=\Omega_{max}}=2.41\,\Msun,
M_{b,max}\vert_{\Omega=\Omega_{max}}=2.84\,\Msun$. There are two
different reasons why such a large-mass model has been chosen. The
first one is that we are interested in the development of a shear
instability in the metastable star produced by the merger of a binary
system of two neutron stars. As shown by a number of authors and most
recently in~\cite{baiotti08}, the product of this merger is either a
supramassive or a hypermassive neutron star. Hence, our reference
model has a mass which is sufficiently large so as to be a reasonable
approximation to the product of a binary neutron star
merger\footnote{We note that the use of a supramassive model has also
  the inconvenient consequence that it is not possible to construct a
  corresponding nonrotating model and this prevents us, for instance,
  from computing the frequency of the fundamental mode and compare it
  with the results of perturbative analyses.}. The second reason is
that a sufficiently massive model is necessary in order to reach
values of the instability parameter which are above the expected
threshold for dynamical bar-mode instabilities as computed
in~\cite{Manca07}, \ie $\beta_{\rm max} \sim 0.25$. Indeed, the
maximum possible value for $\beta$ within the computed sequence is
around $\beta_{\rm max}\simeq 0.2533$ and thus just above the
threshold (\cf long-dashed line in the left panel of figure
\ref{fig:IDb}). Note also that the threshold for the instability tends
to increase for smaller rest-masses.

Overall, when comparing with equilibrium models generated with a
polytropic EOS with $K=100$ and $\Gamma=2$ (see table~1
of~\cite{Baiotti06b}) the realistic EOS models reach higher
compactness (models in~\cite{Baiotti06b} typically have $M/R_e \sim
0.1$) but lower values of $\beta$ (\ie $\beta_{\rm max} \sim 0.2533$
for the models considered here, while $\beta_{\rm max} \sim 0.28$ for
the polytropic models considered in~\cite{Baiotti06b}).

The whole space of parameters is shown in figure~\ref{fig:IDb}, whose
left panel reports the position in the $(\beta, M/R_e)$ plane of the
initial models computed with ${\hat A}=1$, and where the filled
circles represent those we have evolved numerically. Indicated with
solid thin lines are isocontours of constant baryon mass models while
indicated with a thick dashed line is the threshold to the dynamical
bar-mode instability as computed for a $\Gamma=2$
polytrope~\cite{Manca07}. The right panel reports the same initial
models considered in the left one but shown in $(\rho_c, r_p/r_e)$
planes with isocontours of constant $\beta$ (upper part) or constant
$M/R_e$ (lower part). The main properties of the simulated models are
also reported in table~\ref{table:initial}, where we also introduce
our naming convention. Any initial model is indicated as
\texttt{M.\%.\#}, with \texttt{\%} being replaced by the value of the
differential-rotation parameter ${\hat A}$ and \texttt{\#} by the
instability parameter $\beta$. As an example,~\texttt{M.1.200} is the
star with ${\hat A}=1$ and $\beta=0.200$.

Finally, shown in figure~\ref{fig:IDa} are the angular-velocity
profiles (left panel) and the rest-mass density profiles (right panel)
of some representative models, namely~\texttt{M.1.150},
\texttt{M.1.200} and~\texttt{M.1.250}. Indicated with different
symbols, which match the ones in the left panel of
figure~\ref{fig:corotation_band}, are the normalized radial positions
of the corotation radii, with the one for model~\texttt{M.1.150} being
shown filled to help distinguish it from the others. Although we
postpone to sections~\ref{sec:evolution} and ~\ref{sec:necessary
  conditions} the discussion of the implications of these corotation
radii, two aspects of the initial data are worth emphasizing. The
first one is that the amount of differential rotation for a given
value of ${\hat A}$ effectively decreases when increasing the
instability parameter $\beta$ (\cf left panel of
figure~\ref{fig:IDa}), thus resulting in a smaller corotation band for
models with large $\beta$. The second one is that all the initial
models evolved are axisymmetric but have a ``toroidal-topology'',
namely have the maximum density $\rho_{\rm max}$ that is not at the
center of the star, and thus $\rho_{c} < \rho_{\rm max}$. This
toroidal deformation increases with the rotation and thus with
$\beta$.

\begin{figure}[h]
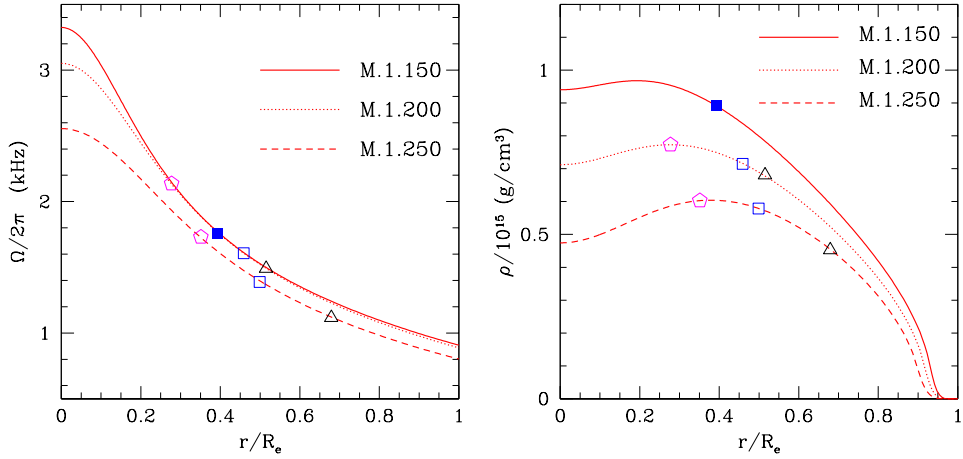

  \begin{center}
    \includegraphics[width=0.495\textwidth]{\imgname{omg_profiles}}
    \includegraphics[width=0.495\textwidth]{\imgname{rho_profiles}}
  \end{center}
  \vglue-.5cm
  \caption{Left panel: Initial angular-velocity profiles for three
    representative models with small, medium and high values of
    $\beta$ and ${\hat A}=1$ (A similar behaviour is shown also by the
    models with ${\hat A}=2$). Indicated with different symbols, which
    match the ones in the left panel of
    figure~\ref{fig:corotation_band} and~\ref{fig:corot_rad}, are the
    normalized radial positions of the corotation radii, with the one
    for model~\texttt{M.1.150} being shown filled to help distinguish
    it. Right panel: the same as in the left panel but for the initial
    rest-mass density.}
  \label{fig:IDa}
\end{figure}

%XXXXXXXXXXXXXXXXXXXXXXXXXXXXXXXXXXXXXXXXXXXXXXXXXXXXXXXXXXXXXXXXXXXXXXXXXX%
%-----------------------------------------------------------------
\section{Methodology of the analysis}
\label{sec:methodology}
%-----------------------------------------------------------------
%XXXXXXXXXXXXXXXXXXXXXXXXXXXXXXXXXXXXXXXXXXXXXXXXXXXXXXXXXXXXXXXXXXXXXXXXXX%

A number of different quantities are calculated during the evolution
to monitor the dynamics of the instability. Among them is the
quadrupole moment of the matter distribution
\begin{equation}
\label{eq:defQuadrupole}
I^{jk} = \int\! d^{3}\!x \; \sqrt{\gamma} W_{_{\!L}} \rho  \; x^{j} x^{k} \,.
\end{equation}
which we compute in terms of the conserved density $\sqrt{\gamma}
W_{_{\!L}} \rho$ rather than of the rest-mass density $\rho$ or of the
$T_{00}$ component of the stress energy momentum tensor.  Of course,
the use of $\sqrt{\gamma} W_{_{\!L}} \rho$ in place of $\rho$ or of
$T_{00}$ is arbitrary and all the three expressions would have the
same Newtonian limit, though with different amplitudes for the
gravitational waveforms produced (see for example
\cite{Baiotti:2008nf}).  However, we here adopt the form
(\ref{eq:defQuadrupole}) because $\sqrt{\gamma}W_{_{\!L}}\rho$ is a
quantity whose conservation is guaranteed by the form chosen for the
hydrodynamics equations.  

The quadrupole moment (\ref{eq:defQuadrupole}) can be conveniently
used to quantify both the growth time of the $m=2$ instability
$\tau_2$ and the oscillation frequency once the instability is fully
developed $\sigma^i_2$. (Hereafter we will indicate respectively with
$\tau_i$ and $\sigma^i_{(n)}$ the growth time and frequencies of the
$m=n$ unstable modes and we note that, as will be discussed later on,
during the simulation a number of different modes appear, thus
justifying the use of the upper index ``$i$''). In practice, we
perform a nonlinear least-square fit of the $xy$ component of the
computed quadrupole $I^{jk}(t)$ and we generally use as fitting
function a sum of $N$ (usually three) exponentially modulated cosines
\begin{equation}  
I^{jk}(t) = \sum_{i=1}^N  I^{jk}_{0(i)} \; \e^{t/\tau_{(i)}} 
               \cos(2\pi\, \sigma^i_{(n)} \, t+\phi_{(i)})\,,
\label{eq:Quadrupolefit}
\end{equation}  
where $I^{jk}_{0} \equiv I^{jk}(t=0)$. Because we commonly have only
about $10$ cycles in the time interval considered for the fits,
extreme care needs to be applied when computing the growth time,
especially when the oscillation frequencies and the growth times are
close to each other. In these cases, in fact, variations of the
initial phase of the modes $\phi_{(i)}$ can result in large variation
of the growth times. In view of this, we will not report them.

Using three components of the quadrupole moment in the $(x, y)$ plane
we can define the distortion parameters $\eta_+(t)$ and
$\eta_\times(t)$, as well as the axisymmetric mode $\eta_0(t)$ as
\begin{equation}
\label{eq:etas}
\fl
\eta_+(t) \equiv \frac{I^{xx}(t)-I^{yy}(t)}{I^{xx}(0)+I^{yy}(0)}\,, 
\quad
\eta_\times(t) \equiv \frac{2I^{xy}(t)}{I^{xx}(0)+I^{yy}(0)}\,, 
\quad
\eta_0(t) \equiv \frac{I^{xx}(t)+I^{yy}(t)}{I^{xx}(0)+I^{yy}(0)}\,, 
\end{equation}
so that the modulus $\eta(t)$ and the instantaneous orientation of the bar
are given by
\begin{equation}
\label{eq:phyQuadrupole}
\fl
\eta(t) =\sqrt{\eta_+(t)^2+\eta_\times(t)^2}  \,,
\qquad
\phi_{\rm bar}(t) = \tan^{-1}\left(\frac{2I^{xy}(t)}{I^{xx}(t)-I^{yy}(t)}\right)\,. 
\end{equation}

Finally,  as a useful tool to describe the nonlinear properties of 
the development and saturation of the instability, the rest-mass density 
is decomposed into its Fourier modes $P_m(t)$ as 
\begin{equation}
{P}_m(t) \equiv\int\!d^3\!x  \, \rho \, \e^{{\rm i} m \phi}\,,
%% \quad {\phi}_m  \, 
\qquad \mbox{where}\quad \phi = \tan^{-1} (x/y). 
\label{eq:modes}
\label{eq:phimodes}
\end{equation}
The phase ${\phi}_m\equiv \arg ({P}_m)$ essentially provides the
instantaneous orientation of the $m$-th mode when the corresponding
mode has a nonzero power.  Note that despite their denomination, the
Fourier modes (\ref{eq:modes}) do not represent proper eigenmodes of
oscillation of the star. While, in fact, the latter are well defined
only within a perturbative regime, the former simply represent a tool
to quantify, within the fully nonlinear regime, what are the main
components of the rest-mass distribution. As a final comment we note
that while all quantities
(\ref{eq:defQuadrupole})--(\ref{eq:phimodes}) are expressed in terms
of the coordinate time $t$ and are not invariant measurements, the
lengthscale of variation of the lapse function at any given time is
always larger than twice the stellar radius at that time, ensuring
that events on the same timeslice are also close in proper time.

%-----------------------------------------------------------------
\section{Results}
\label{sec:results}
%-----------------------------------------------------------------

In what follows we first describe the dynamics of the shear
instability as deduced from the numerical simulations and then
contrast our results with the phenomenological predictions on the
necessary conditions for its development.

%-----------------------------------------------------------------
\subsection{Dynamics of the instability}
\label{sec:evolution}
%-----------------------------------------------------------------

\begin{figure}[t]
\begin{center}
\includegraphics[width=0.425\textwidth]{\imgname{M.1.150.dynamics}}
\hskip 0.5cm 
\includegraphics[width=0.425\textwidth]{\imgname{M.1.160.dynamics}}
\includegraphics[width=0.425\textwidth]{\imgname{M.1.200.dynamics}}
\hskip 0.5cm 
\includegraphics[width=0.425\textwidth]{\imgname{M.1.220.dynamics}}
\includegraphics[width=0.425\textwidth]{\imgname{M.1.241.dynamics}}
\hskip 0.5cm 
\includegraphics[width=0.425\textwidth]{\imgname{M.1.250.dynamics}}
\end{center}
\vglue-0.5cm
\caption{Summary of the dynamics of some representative models, \ie
 ~\texttt{M.1.150, M.1.160, M.1.200, M.1.220, M.1.241} and
 ~\texttt{M.1.250}, with increasing values of $\beta$. For each panel
  the upper part reports the evolution of the bar-distortion parameter
  $\eta_+$, while the lower part shows the evolution of the power in
  the different modes of the Fourier decomposition of the rest-mass
  density $P_m$. See text for details. Note that we here report
  only some models and with ${\hat A}=1$ as they are representative
  also of those with ${\hat A}=2$.}
%%%%%%%%%%%%%%%%%%%%%%%%%%%%%%%%%%%%%%%%
\label{fig:evolutionGENERAL}
\end{figure}

\begin{figure}[t]
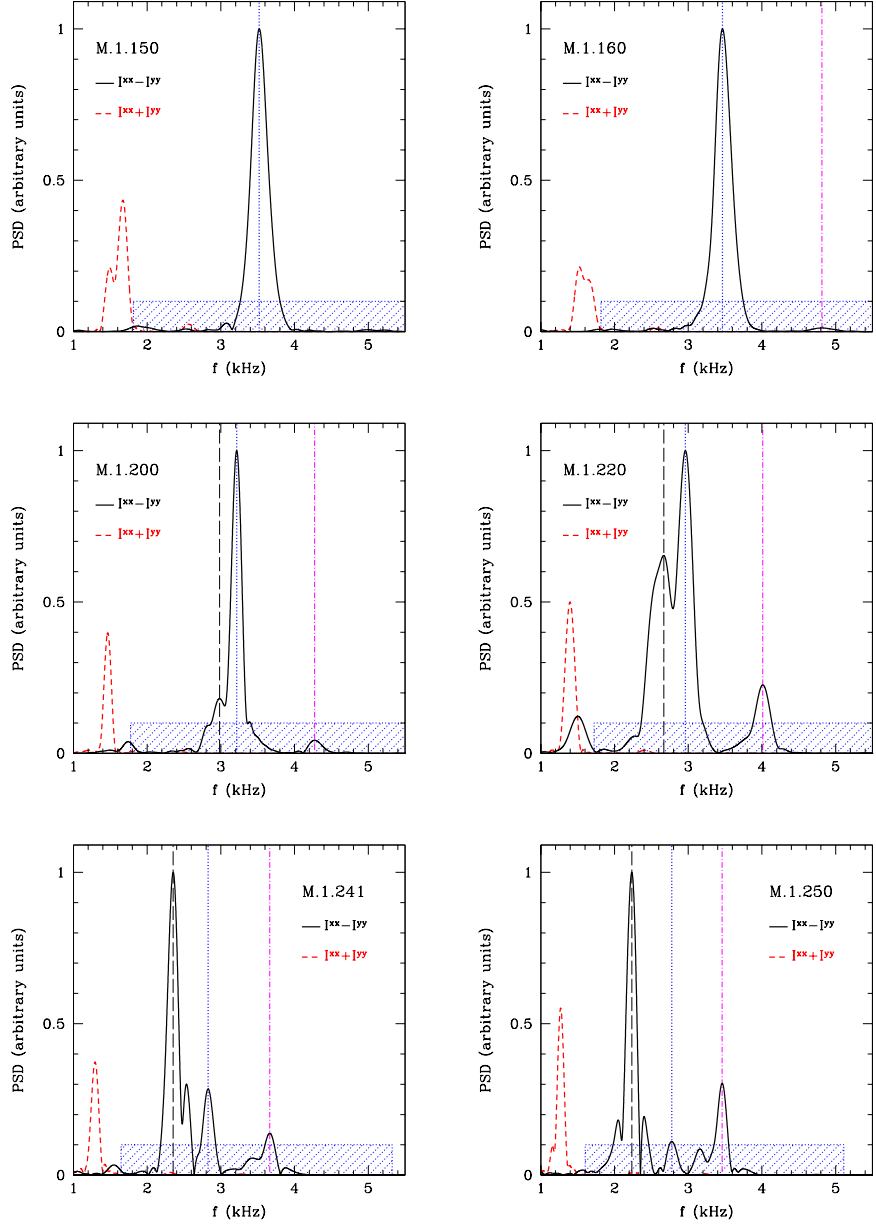

\begin{center}
\includegraphics[width=0.425\textwidth]{\imgname{M.1.150.SPECdynamics}}
\hskip 0.5cm 
\includegraphics[width=0.425\textwidth]{\imgname{M.1.160.SPECdynamics}}
\includegraphics[width=0.425\textwidth]{\imgname{M.1.200.SPECdynamics}}
\hskip 0.5cm 
\includegraphics[width=0.425\textwidth]{\imgname{M.1.220.SPECdynamics}}
\includegraphics[width=0.425\textwidth]{\imgname{M.1.241.SPECdynamics}}
\hskip 0.5cm 
\includegraphics[width=0.425\textwidth]{\imgname{M.1.250.SPECdynamics}}
\end{center}
\vglue-0.5cm
\caption{Power spectral density (filtered using Hanning windowing) in
  arbitrary units of the evolution of $\eta_+$ for the models
  described in figure~\ref{fig:evolutionGENERAL} (black continuous
  line) and of the axisymmetric mode $\eta_0$ (red dashed
  line). Indicated with vertical lines (long-dashed black, dotted blue 
   and dot-dashed magenta) are the peak frequencies reported in the left panel of
  figure~\ref{fig:corotation_band} within the corotation band and
  which are used in figure~\ref{fig:corot_rad} to mark the position of
  the corotation radii. Finally, shown with a shaded rectangular area
  is the corotation band for all models.}
\label{fig:PSD}
\end{figure}

As mentioned when discussing the initial data, we have evolved
numerically two sequences of constant baryonic mass, with $M_b=2.5\,
M_\odot$. The first sequence has a higher degree of differential
rotation (\ie ${\hat A}=1$) and instability parameter ranging from
$\beta=0.140$ to $\beta=0.250$. The second sequence, on the other
hand, has a smaller degree of differential rotation (\ie $\hat{A}=2$)
and instability parameter between $\beta=0.125$ and $\beta=0.200$. For
both sequences the highest value of $\beta$ considered is also very
close to the highest attainable with our initial data code (see figure
\ref{fig:IDb}). To keep the computational costs to an affordable level
we have evolved of all these models for $\sim 15\,\mss$ and although
this time window is in general insufficient to capture the suppression
of the instability, is adequate to measure the frequencies of the
unstable modes and provide a first estimate of the growth times.

After analyzing the results following the method outlined in the
previous section and focusing on the properties of the distortion
parameters $\eta_+$, $\eta_\times$ and $\eta_0$, we find that
\textit{all of the models} show an $m=2$ instability. The maximum
distortions obtained during the simulation time are $\eta \sim 0.1$
(\ie distortions of $~10\%$ with respect to the axisymmetric
progenitor), with the maximum values being reached for the models of
the ${\hat A}=1$ sequence. The models of the ${\hat A}=2$ sequence, in
fact, have in general much smaller distortions, with $\eta \sim
0.01$ over the timescale over which they have been evolved. We believe
this is simply the consequence of the fact that the models in this
sequence have smaller growth rates (see also the discussion in the
next section).

To show the general behaviour of the simulation we focus on the
description of the result of the simulation for the ${\hat A}=1$
sequence as this is representative also of the one with ${\hat A}=2$.
In particular, in the upper part of each panel in figure
\ref{fig:evolutionGENERAL} we show the time evolution of the
distortion parameter $\eta_{+}$ computed with Eq.~(\ref{eq:etas}) for
six representative models:~\texttt{M.1.150},~\texttt{M.1.160},
\texttt{M.1.200},~\texttt{M.1.220},~\texttt{M.1.241} and
\texttt{M.1.250}. Similarly, in the lower parts of each panel we show
the corresponding evolution of the Fourier-modes computed from
Eq.~(\ref{eq:modes}) for $m=1,2,3,4$.

Note that for models with lower values of $\beta$ the bar-mode
deformation is very similar to the one already discussed
in~\cite{Baiotti06b,Manca07}, growing exponentially and with only one
unstable mode appearing (\cf first row of figure
\ref{fig:evolutionGENERAL}). However, as the rotation is increased,
the development of the instability is more complex and at least two
unstable modes appear which develop in different parts of the
star. These two modes have very similar but distinct frequencies and
growth times, leading to a series of beatings in the evolutions of
$\eta_{+}$, whose irregular evolution makes the calculation of the
growth times challenging (\cf second row). As the rotation is
increased towards the maximal values of $\beta$, up to three distinct
modes appear and the evolution of the instability is correspondingly
more complex (\cf third row). A similar behaviour is shown also by the
evolution of the Fourier modes. The two models with lower $\beta$, in
fact, show a clear growth of the $m=2$ mode and of the $m=1$, with the
latter becoming first comparable and then larger when the instability
is suppressed. In models~\texttt{M.1.250, M.1.241}, on the other hand,
the $m=2$ and $m=3$ modes have comparable amplitude for a long period
and then the $m=2$ becomes the dominant one (this is more clear in
model~\texttt{M.1.241}). Finally, for larger values of $\beta$, the
$m=1$ mode never attains values comparable with either the $m=2$ or
the $m=3$, which instead control the evolution.

Figure~\ref{fig:PSD} reports the power spectral density (PSD) of
$\eta_+$ and $\eta_0$ for the same $6$ models (solid black lines) and
allows to appreciate how the spectrum changes as the instability
parameter is increased. More specifically, it is very apparent that at
low $\beta$ the spectra have only one peak, whose maximum is marked by
the vertical blue dotted lines and that is present at all the values
of beta considered; this is the mode we refer to as $\sigma^2_2$ (\cf
table~\ref{table:results} and figure~\ref{fig:corotation_band}). As
$\beta$ increases, the amplitude of this peak decreases and it becomes
the weakest for very large rotation rates. Starting from
model~\texttt{M.1.200} (although a hint of a peak is present already
in~\texttt{M.1.160}), a second peak appears at higher frequency and is
marked with magenta vertical dot-dashed lines; this is the mode we
refer to as $\sigma^3_2$ and although it never becomes the largest
one, its amplitude increases with $\beta$. Finally, a third peak
appears at low frequency in the spectra starting from
model~\texttt{M.1.220} (even though a hint is present also for
model~\texttt{M.1.210}) and is marked with black vertical long-dashed
lines. This mode, which we refer to as $\sigma^1_2$, becomes the
dominant one at high $\beta$. Note also that in each panel of
figure~\ref{fig:PSD} we report with a dashed blue box the corotation
band that will be further discussed in the following section.

As a final remark we note that the peaks in the spectrum of $I^{xx}+I^{yy}$
(red dashed lines in figure~\ref{fig:PSD}) are related to axisymmetric
$m=0$ modes and most likely to $f$-modes oscillation excited by the
development of the instability. Indeed, their frequencies
$\sigma_{_f}$ match reasonably well the phenomenological fit for the
$f$-mode frequencies of nonrotating neutron stars with realistic EOSs
computed in~\cite{Benhar:2004xg}. Because our sequences do not contain
nonrotating configurations (the rest mass is larger than the maximum
one), this association is just qualitative and a more detailed
investigation of the frequency spectra of the equilibrium models is
necessary to confirm this suggestion.

\begin{table}[t]
\caption{Results of the analysis of quadrupole evolutions. The
  frequencies $\sigma_{_f}$ and $\sigma_2^{i}$ are obtained from the
  position of the peaks in the PSD while the frequencies
  $\tilde{\sigma}_2^{i}$ are obtained from the nonlinear fitting of
  Eq.~(\ref{eq:Quadrupolefit}).  Reported in the last two columns are
  the edges of the corotation band as expressed in terms of angular
  frequency at the surface $\Omega_e/\pi\equiv\Omega(r=R_e)/\pi$ and
  on the axis $\Omega_c/\pi$. Note that depending on the rate of
  rotation and the degree of differential rotation some frequencies
  may not be present.}
%%%%%%%%%%%%%%%%%%%%%%%%%%%%%%%%%%%%%
\label{table:results}
\begin{indented}
\item[]
%%\hspace{-2.5cm}
\hspace{-1.0cm}
\begin{tabular}{@{}cccccccccc}
\br
Model & $\sigma_{f}$ 
      & $\sigma_2^{1}$ & $\sigma_2^{2}$ & $\sigma_2^{3}$ 
      & $\tilde{\sigma}_2^{1}$ & $\tilde{\sigma}_2^{2}$ & $\tilde{\sigma}_2^{3}$ 
      & $\Omega_e/\pi$ & $\Omega_c/\pi$ \\[1mm]
      & (kHz) & (kHz) & (kHz) & (kHz) & (kHz) & (kHz) & (kHz) & (kHz) & (kHz)\\ [1mm] 
\hline
\texttt{M.1.140} & $1.678$ &     $-$ & $3.601$ & $-$     & $-$     & $3.521$ & $-$     & $1.807$ & $6.701$ \\
\texttt{M.1.150} & $1.671$ &     $-$ & $3.521$ & $-$     & $-$     & $3.515$ & $-$     & $1.814$ & $6.649$ \\
\texttt{M.1.160} & $1.524$ &     $-$ & $3.465$ & $4.815$ & $-$     & $3.428$ & $4.826$ & $1.816$ & $6.577$ \\
\texttt{M.1.180} & $1.515$ &     $-$ & $3.318$ & $4.502$ & $3.200$ & $3.335$ & $4.522$ & $1.805$ & $6.376$ \\
\texttt{M.1.200} & $1.466$ & $2.983$ & $3.215$ & $4.272$ & $2.719$ & $3.267$ & $4.249$ & $1.774$ & $6.102$ \\
\texttt{M.1.210} & $1.432$ & $2.796$ & $3.220$ & $4.152$ & $2.846$ & $3.174$ & $4.180$ & $1.751$ & $5.940$ \\
\texttt{M.1.220} & $1.396$ & $2.668$ & $2.962$ & $4.012$ & $2.559$ & $2.956$ & $4.022$ & $1.723$ & $5.760$ \\
\texttt{M.1.230} & $1.366$ & $2.488$ & $2.920$ & $3.871$ & $2.449$ & $2.900$ & $3.843$ & $1.689$ & $5.563$ \\
\texttt{M.1.241} & $1.293$ & $2.353$ & $2.828$ & $3.664$ & $2.342$ & $2.810$ & $3.676$ & $1.645$ & $5.324$ \\
\texttt{M.1.244} & $1.297$ & $2.328$ & $2.788$ & $3.632$ & $2.301$ & $2.773$ & $3.603$ & $1.632$ & $5.256$ \\
\texttt{M.1.247} & $1.274$ & $2.267$ & $2.750$ & $3.564$ & $2.250$ & $2.765$ & $3.522$ & $1.618$ & $5.183$ \\
\texttt{M.1.250} & $1.270$ & $2.235$ & $2.777$ & $3.461$ & $2.224$ & $2.768$ & $3.475$ & $1.603$ & $5.110$ \\
\hline             
\hline             
\texttt{M.2.125} & $1.736$ & $2.969$ & $3.384$ & $-$     & $2.964$ & $3.395$ & $-$     & $2.217$ & $3.929$ \\
\texttt{M.2.150} & $1.682$ & $3.033$ & $3.322$ & $-$     & $3.050$ & $3.313$ & $-$     & $2.232$ & $3.868$ \\
\texttt{M.2.175} & $1.610$ & $2.901$ & $3.146$ & $-$     & $2.886$ & $3.030$ & $-$     & $2.222$ & $3.766$ \\
\texttt{M.2.200} & $1.531$ & $2.589$ & $2.910$ & $-$     & $2.512$ & $2.742$ & $-$     & $2.125$ & $3.484$ \\
\br
\end{tabular}
\end{indented}
\end{table}

%-----------------------------------------------------------------
\subsection{Necessary conditions for the instability}
\label{sec:necessary conditions}
%-----------------------------------------------------------------

To support the interpretation of these instabilities as shear
instabilities we have considered whether the necessary conditions
suggested by~\cite{Watts:2003nn} and by~\cite{Ou06} are met. We recall
that Watts and collaborators pointed out that an unstable
configuration should have the unstable mode with a frequency within
the corotation band. In the case of the Newtonian
expression~(\ref{eq:velocityProfile_newt}), the corotation band is
simpler to compute and for a mode with frequency $\sigma$ and
azimuthal number $m$, this is simply given by
\begin{equation}
\frac{\Omega_c {\hat A}^2}{{\hat A}^2+1} < \frac{\sigma}{m} 
<\Omega_c\,.
\end{equation}
We have therefore checked whether any of the unstable $m=2$
(bar-modes) with computed frequency $\sigma^i_{(2)}$ has pattern speed
velocity, $\sigma^i_2/m=\sigma^i_2/2$, within the corotation band. A
careful and rather involved analysis has indeed confirmed the
prediction of Watts and collaborators: namely, \textit{all the
  unstable modes are within the corotation band} of the progenitor
axisymmetric model. These results for the models with ${\hat A}=1$ are
listed in table~\ref{table:results}, where we report the edges of
the general-relativistic corotation band in terms of angular frequency
at the surface $\Omega_e/\pi\equiv\Omega(r=R_e)/\pi$ and on the axis
$\Omega_c/\pi$, as well as the frequencies of the unstable $m=2$ modes
obtained from the position of the peaks in the PSD (\ie
$\sigma_2^{i}$) or from the nonlinear fitting of
Eq.~(\ref{eq:Quadrupolefit}) (\ie $\tilde{\sigma}_2^{i}$). Note that
depending on the rate of rotation and the degree of differential
rotation some frequencies may not be present in the corotation band.

\begin{figure}[h]
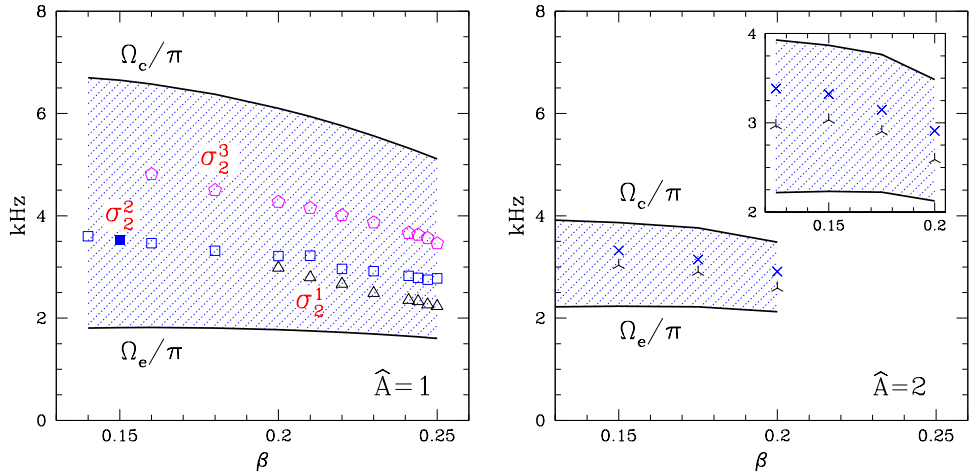

\begin{center}
\includegraphics[width=0.495\textwidth]{\imgname{freqs_vs_beta}}
\includegraphics[width=0.495\textwidth]{\imgname{freqs_vs_betaA2}}
\end{center}
\vglue-.5cm
\caption{Left panel: Position of the non-axisymmetric frequencies
  $\sigma^i_2$ for models with ${\hat A}=1$ and shown as a function of
  $\beta$; indicated with thick solid lines are the edges of the
  corotation band, \ie the frequency interval between $\Omega_c/\pi$
  and $\Omega_e/\pi$. Shown with a filled symbol, to help distinguish
  it from the others, is the $\sigma^1_2$ frequency for model
  ~\texttt{M.1.150} (\cf figure~\ref{fig:IDa}
  and~\ref{fig:corot_rad}). Right panel: the same as in the left one
  and with the same scale but for models with ${\hat A}=2$; the inset
  shows instead a magnified view. Clearly, in both panels all the
  unstable modes are within the band and up to three unstable modes
  appear for the ${\hat A}=1$ sequence with increasing $\beta$.}
\label{fig:corotation_band}
\end{figure}

\begin{figure}[t]
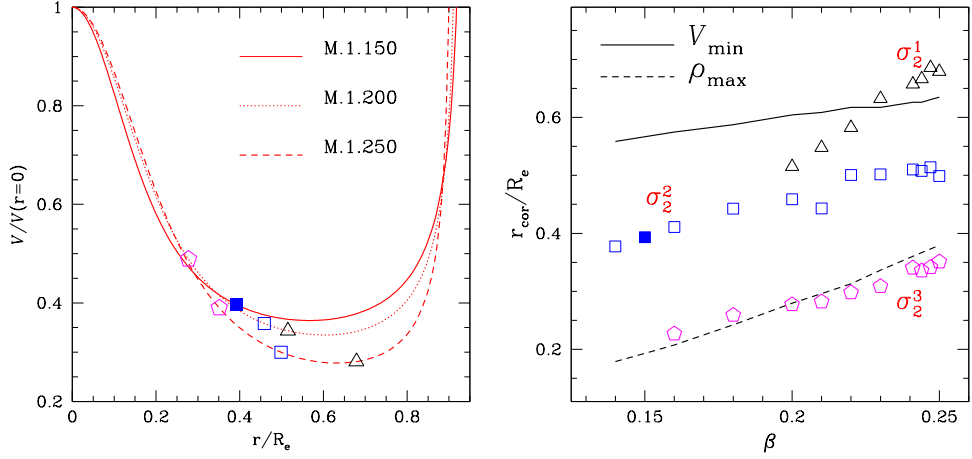

\begin{center}
\includegraphics[width=0.495\textwidth]{\imgname{vortensity_profiles}}
\includegraphics[width=0.495\textwidth]{\imgname{vortensity_min_radii}}
\end{center}
\vglue-.5cm
\caption{Left panel: radial profiles of the vortensity ${\cal V}$ for
  three representative models with small, medium and high values of
  $\beta$. The different symbols match the ones in
  figure~\ref{fig:IDa} and in the left panel of
  figure~\ref{fig:corotation_band}, with the one for model
 ~\texttt{M.1.150} reported as filled, and show the actual position of
  the corotation radii. Note that none of these coincides with the
  minimum of ${\cal V}$; for compactness we have reported only the
  models with ${\hat A}=1$. Right panel: normalized corotation radii
  for the different frequencies $\sigma^i_2$ presented in the left
  panel of figure~\ref{fig:corotation_band} shown as a function of
  $\beta$; the same convention of the previous figures is used for the
  different symbols.}
\label{fig:corot_rad}
\end{figure}

The data in table~\ref{table:results} is also shown in the left panel
of figure~\ref{fig:corotation_band}, where we plot the position of the
non-axisymmetric frequencies $\sigma^i_2$ within the corotation band
as a function of $\beta$; indicated with a filled symbol, to
distinguish it from the others, is the $\sigma_1$ frequency for model
\texttt{M.1.150} (\cf figure~\ref{fig:IDa} and~\ref{fig:corot_rad}).
The right panel of the same figure shows the corresponding information
and on the same scale but for the models with ${\hat A}=2$ (the inset
shows instead a magnified view). Clearly, for both sequences all the
unstable modes are within the band and, as the stellar rotation rate
increases, more unstable modes appear for the same value of $\beta$.

Watts, Andersson and Jones also give a qualitative description,
summarized in their figure 2 of~\cite{Watts:2003nn}, of how the
unstable and stable models should be distributed in the ($\hat{A},
\beta$) plane. The considerations they make are particularly simple:
for high degrees of differential rotation, that is at low values of
$\hat{A}$, the corotation band is rather wide and there will be both
an upper value and a lower value of $\beta$ between which the shear
instability can develop (these critical values of $\beta$ correspond
to the entrance and exit of the unstable mode in the corotation
band). This situation corresponds therefore to the one commonly
encountered in numerical simulations, such as those
in~\cite{Shibata:2003yj,Ott:2005gj, Ou06, Cerda07b, Ott07a,
  Ott07b,scheidegger08}, and for which the shear instability takes
place only for very small values of the instability parameter and on
timescales that are much longer than the dynamical one. When moving to
larger degrees of differential rotation, that is when going to smaller
values of $\hat{A}$, the corotation band becomes larger and larger and
the shear instability can develop essentially for all values of
$\beta$, merging with the dynamical bar-mode instability for $\beta
\gtrsim 0.25$. This is exactly what has been found here. Conversely,
when moving to smaller degrees of differential rotation, that is when
going to higher values of $\hat{A}$, the corotation band becomes
thinner and the shear instability can develop only for a smaller range
of $\beta$. As the differential rotation is further decreased and the
star tends to rotate uniformly, the corotation band width vanishes,
all the models are stable to the shear instability and subject only to
the dynamical bar-mode instability for $\beta \gtrsim 0.25$. This is
indeed the case for the unstable models evolved in~\cite{Baiotti06b},
which were purely (bar-mode) dynamically unstable and none of which
had the unstable mode within the corresponding corotation band, but
above it.

In essence, therefore, there should be an intermediate range of
$\hat{A}$ for which the instability is absent at low $\beta$, appears
at intermediate values and then disappears again at high $\beta$, thus
defining an interval of values of $\beta$ for which the models are
unstable. To validate also this prediction we simulated a sequence of
models with smaller differential rotation and $\hat{A}=2$. This
seconds sequence has the same baryonic mass as that with $\hat{A}=1$,
but with $\beta$ in a smaller range, namely between $\beta=0.125$ and
$\beta=0.200$ (Note that $\beta=0.200$ is also very close to the
largest value for which we could build an equilibrium model.)
Unfortunately, also all of these models show an $m=2$ shear
instability, with the unstable frequencies falling within the
corotation band, as it was for the $\hat{A}=1$ stars. Of course, lack
of evidence is not evidence of absence and the fact that we have not
found stable models within our range of values of $\beta$ most likely
means that these stable models have to be searched either for values
of $\beta < 0.125$ (for this value of $\hat{A}=2$), or by moving
towards higher values of $\hat{A}$, where the corotation band is less
large.

An obvious consequence of the phenomenological scenario described
by~\cite{Watts:2003nn} and confirmed by the calculations reported here
is that the idea of a low-$T/|W|$ instability is indeed misleading
and it is instead more meaningful to think of a more generic shear
instability that, depending on the degree of rotation and of
differential rotation, may manifest itself on timescales that are
comparable with the dynamical ones (as in the cases reported here) or
on much longer ones (as in the cases reported
in~\cite{Shibata:2002mr}).

Another phenomenological prediction made by Watts, Andersson and Jones
in~\cite{Watts:2003nn}, is that the growth times should be shorter in
the center of the band and increase towards the edges (the growth
times should in fact diverge at the edges). As shown in
figure~\ref{fig:corotation_band}, all of the simulated models do have
unstable modes well inside the corotation band and this probably
explains why we observe them develop on a timescale which is
comparable with the dynamical one and not on much longer timescales as
in the original finding of~\cite{Shibata:2002mr}. We also find
indications that the $\hat{A}=2$ models, which generically have longer
growth times, have unstable modes which occupy regions of the
corotation band that are overall more central (\cf right panel of
figure~\ref{fig:corotation_band}) and thus in contrast with what
expected from~\cite{Watts:2003nn}. However, the difficulties mentioned
above in computing an accurate estimate of the growth times and the
intrinsic difficulties in determining what is the central part of the
corotation band, prevent us from providing a more quantitative
validation of this prediction.

We now switch to consider our results in the light of the other
necessary condition for the onset of the shear instability discussed
in~\cite{Ou06}. As mentioned in the Introduction, it has in fact been
shown that the presence of a minimum in the profile of the vortensity
is a necessary condition for a mode in corotation to be
unstable~\cite{Ou06}. The intuitive description is that the vortensity
well can act as a resonant cavity inside the star, amplifying the
modes that happen to lay near its minimum~\cite{Papaloizou89}. Indeed,
the growth rate of the unstable mode is expected to depend on the
location of its corotation radius with respect to the vortensity
profile, being proportional to the depth of its corotation radius
inside the vortensity well~\cite{Ou06}. Note that the mere existence
of a local minimum in the vortensity cannot be used as a sufficient
condition for the occurrence of a shear instability. All of the
unstable models considered in~\cite{Baiotti06b}, in fact, do show a
local minimum in the vortensity but do not have the unstable modes in
the corotation band.

To validate whether this Newtonian condition holds also for the
general-relativistic instabilities simulated here, we have computed
for all the models the Newtonian vortensity, which is defined as the
ratio, along the radial cylindrical coordinate, between the radial
vorticity and the density
\begin{equation}
\mathcal V = \frac{2\Omega+\varpi\Omega_{,\bar{\varpi}}}{\rho},
\end{equation}
where $\varpi$ is the radial cylindrical coordinate (a fully
general-relativistic definition of the vortensity is also possible but
more complicated to compute and not significantly different from the
Newtonian one). When doing this we found that all the unstable models
have vortensities with a local minimum in the star and this is shown
in the left panel of figure \ref{fig:corot_rad}, which reports the
radial profiles of the vortensity ${\cal V}$ for three representative
models with small, medium and high values of $\beta$. The different
symbols match the ones in the left panel of figure~\ref{fig:IDa}
and~\ref{fig:corotation_band}, with the one for model~\texttt{M.1.150}
reported as filled, and show the actual positions of the corotation
radii. The right panel of the same figure, on the other hand, shows
the normalized corotation radii for the different frequencies
$\sigma^i_2$ presented in the left panel of
figure~\ref{fig:corotation_band} as a function of $\beta$.

Note that none of these coincides systematically with the minimum of
${\cal V}$, nor with the maximum of the rest-mass density. All of the
corotation radii, however, do move towards larger radial positions as
the rotation rate is increased, exactly as does the minimum of the
vortensity and the rest-mass maximum. Although we cannot confirm that
modes whose corotation radius is closer to the minimum of the
vortensity have systematically shorter growth times (our data is not
sufficiently accurate for this), we can compare the models with the
higher degree of differential rotation ${\hat A}=1$ with those having
a smaller degree of differential rotation ${\hat A}=2$. In the first
case the vortensity well is considerably deeper, with $\Delta
\mathcal{V}/\mathcal{V} = 1 - \mathcal{V}_{\rm min}/\mathcal{V}_{r=0}
\sim 0.6-0.7$, while for the latter the vortensity has shallower wells
with $\Delta \mathcal{V}/\mathcal{V} = 1 - \mathcal{V}_{\rm
  min}/\mathcal{V}_{r=0} \sim 0.02-0.25$. Since the models with ${\hat
  A}=2$ have smaller growth rates, our results indicate therefore that
also the existence of a local minimum in the vortensity can be taken
as a necessary condition for the development of the shear instability
and that the depth of the vortensity well can be used to estimate the
growth of the instability.

%XXXXXXXXXXXXXXXXXXXXXXXXXXXXXXXXXXXXXXXXXXXXXXXXXXXXXXXXXX%
%-----------------------------------------------------------------
\section{Conclusions}
\label{sec:conclusions}
%-----------------------------------------------------------------

For many years the properties of rapidly rotating and self-gravitating
fluids have been characterized by a complete analytic perturbative
theory which provided, for instance, sufficient conditions for the
development of instabilities (see, for instance,
~\cite{Chandrasekhar69c,Lai93} for a collection of results). As
numerical simulations have become increasingly accurate and stable on
longer timescales, these predictions, both in Newtonian theory and in
full general relativity, have been verified, corrected and in some
cases extended. Our ability of modelling such configurations has now
reached a maturity such that a number of new properties and
instabilities have been ``discovered'' numerically, but which do not
have behind a fully perturbative description. The main reason for this
is that most of this phenomenology is the result of physical scenarios
which are much more complex than the ones investigated perturbatively
in the past, \eg non-isolated systems with high differential rotation
and exchanging mass and angular momentum, which are much more
difficult to treat analytically.

A most notable example of these complex and yet ubiquitous
instabilities is the so-called ``low-$T/|W|$ instability'', which was
initially found in~\cite{Shibata:2002mr} and then reproduced in a
number of other different scenarios~\cite{Shibata:2003yj,Ott:2005gj,
  Ou06, Cerda07b, Ott07a, Ott07b,scheidegger08,Abdikamalov:2009aq}.
As the phenomenological description of this instability provided by
the simulations has become richer and richer, a full understanding of
the mechanisms that lead to its development has lagged behind and it
is presently unclear. We are therefore in a situation in which
numerical simulations can probe regimes and conditions which are not
yet accessible to perturbative calculations, and can guide the latter
by confirming or refuting those scenarios that although possible do
not find a realization in practice.

This work has followed this spirit and has used fully
general-relativistic calculations of rapidly and differentially
rotating neutron stars modeled with a realistic EOS to shed some light
on the development of the low-$T/|W|$ instability. In particular we
have concentrated our attention on validating an indirect prediction,
made by Watts, Andersson and Jones~\cite{Watts:2003nn}, who recognized
the low-$T/|W|$ instability as the manifestation of a more generic
class of instabilities associated to the existence of a corotation
band~\cite{Balbinski85b,Luyten:1990}, the \textit{shear
  instabilities}, and that should develop for \textit{any value} of the
instability parameter $\beta$ when sufficient amounts of differential
rotation are present. This is exactly what we have found in our
simulations. More specifically, we have performed simulations of
sequences of neutron-star models described by a realistic SLy
EOS~\cite{Douchin01} and having constant rest-mass and degrees of
differential rotation, but with different amounts of rotation. In all
cases considered we have found the development of a bar-mode
instability growing on a dynamical timescale, even when the initial
axisymmetric models were well below the critical limit for the
dynamical bar-mode instability. These results, which match well the
phenomenological scenario portrayed in~\cite{Watts:2003nn}, suggest
therefore that the idea of a low-$T/|W|$ instability is indeed
misleading and should be replaced by the more general one of shear
instability. Depending then on the degree of rotation and of
differential rotation, the instability will develop on timescales that
are comparable to the dynamical one (as reported here) or on much
longer ones (as reported in the first low-$T/|W|$ instability
studies).

Special attention has also been paid to the properties of the unstable
modes and to their position within the corotation band or the
vortensity profiles. In particular, we have shown that all the
unstable modes are within the corotation band of the progenitor
axisymmetric model (which is the necessary condition for the
development of the instability proposed by~\cite{Watts:2003nn}) and
that all of the unstable models have vortensity profiles with a local
minimum (which is the necessary condition suggested
by~\cite{Ou06}). Finally, by comparing the growth times among models
with different degree of differential rotation we have shown that
there is a correlation, although not a strong one, between the depth
of the vortensity well and growth rate of the instability, with the
latter being larger for models with deeper wells.

In summary, the results presented here shed some light on several
aspects of shear instabilities, but they also reveal that more work is
required, for instance, to distinguish between the predictions based
on the corotation band and the ones based on the vortensity well, or
to establish whether in effect they just represent two different ways
of expressing the same physical conditions. Clarifying these aspects
will requires additional analytical and numerical modelling, and this
will be part of our future research.

%XXXXXXXXXXXXXXXXXXXXXXXXXXXXXXXXXXXXXXXXXXXXXXXXXXXXXXXXXXX%

\ack 

It is a pleasure to thank Anna Watts for enlightening discussions and
for sharing her notes with us, as well as Nils Andersson, Ian Jones,
Joel Tohline for useful comments. We are grateful to Nikolaous
Stergioulas for the code to compute the initial data used in these
calculations and to E.~Schnetter and all the \texttt{Carpet/Cactus}
developers. The computations were mainly performed on the Damiana
cluster at the AEI and, in a part, on the Parma's ALBERT2 cluster and
on the bcx cluster at CINECA. This work was supported in part by the
DFG grant SFB/Transregio~7 ``Gravitational-Wave Astronomy'', by
``CompStar'', a Research Networking Programme of the European Science
Foundation, and by the ''Della Riccia'' Foundation.

\section*{References}
%\bibliographystyle{iopart-num}
%\bibliography{aeireferences}
\section*{References}
%bibliographystyle{iopart-num}
%bibliography{aeireferences}
%end{document}

\providecommand{\newblock}{}

\end{document}